\documentclass[preprint,superscriptaddress,showpacs,amsmath,amssymb,aps,pra]{revtex4-2}
\usepackage{pifont}
\usepackage{times}
\usepackage{amssymb}
\usepackage{graphicx}
\usepackage{dcolumn}
\usepackage{dsfont}
\usepackage{bm}
\usepackage{subfigure}
\usepackage[ruled]{algorithm2e}
\usepackage{hyperref}
\usepackage{appendix}
\usepackage{lipsum}

\hypersetup{colorlinks=true, citecolor=blue, urlcolor=blue, linkcolor=blue}

\newtheorem{theorem}{Theorem}

\newtheorem{lemma}{Lemma}

\input amssym.def

\begin{document}
	\title{Bounds on positive operator-valued measure based coherence of superposition}
	\author{Meng-Li Guo}
	\affiliation{School of Mathematical Sciences, Capital Normal University, Beijing 100048, China}
	\author{Jin-Min Liang}
	\affiliation{School of Mathematical Sciences, Capital Normal University, Beijing 100048, China}
	\author{Bo Li}
	\affiliation{School of Computer and Computing Science, Hangzhou City University, Hangzhou 310015, China}
	\author{Shao-Ming Fei}
	\email{feishm@cnu.edu.cn}
	\affiliation{School of Mathematical Sciences, Capital Normal University, Beijing 100048, China}
	\affiliation{Max–Planck-Institute for Mathematics in the Sciences, Leipzig 04103, Germany}
	\author{Zhi-Xi Wang}
	\email{wangzhx@cnu.edu.cn}
	\affiliation{School of Mathematical Sciences, Capital Normal University, Beijing 100048, China}
	
\begin{abstract}
Quantum coherence is a fundamental feature of quantum physics and plays a significant role in quantum information processing. By generalizing the resource theory of coherence from von Neumann measurements to positive operator-valued measures (POVMs), POVM-based coherence measures have been proposed with respect to the relative entropy of coherence, the $l_1$ norm of coherence, the robustness of coherence and the Tsallis relative entropy of coherence. We derive analytically the lower and upper bounds on these POVM-based coherence of an arbitrary given superposed pure state in terms of the POVM-based coherence of the states in superposition. Our results can be used to estimate range of quantum coherence of superposed states. Detailed examples are presented to verify our analytical bounds.
\end{abstract}

\maketitle

\section{Introduction}
Originated from the superposition principle of quantum mechanics, quantum coherence and entanglement are important quantum resources in quantum information processing and quantum computation \cite{Jozsa2003,Horodecki2009,Designolle2021Set}. However, in general the entanglement of a superposed pure state cannot be simply expressed as a linear summation of the entanglement of the individual states in the superposition. Linden \emph{et al}. first investigated the relations between the entanglement of a superposed pure state and the entanglement of the states in the superposition \cite{Linden2006} and Gour \emph{et al}. provided the upper and lower bounds on superposed entanglement based on the Von Neumann entropy of the reduced states \cite{Gour2007} and the entanglement measure concurrence \cite{Niset2007,Con2011}.

Baumgratz, Cramer and Plenio \cite{2014prl} first proposed the resource theory of coherence, established a rigorous coherence quantification framework (BCP framework), and identified computable measures of coherence. Based on the BCP framework, coherence has witnessed theoretically and experimentally fruitful progress \cite{Plenio2017,Fan2018,Winter2016,Bu2017,Tsallis2018,Xiong2018,Guo2020}. For a quantum system associated with a $d$-dimensional Hilbert space $H$, the BCP framework takes into account the coherence defined by an orthogonal basis $\{|j\rangle \}_{j=1}^{d}$, which we call standard coherence. A standard orthogonal basis $\{|j\rangle \}_{j=1}^{d}$ corresponds to a rank-$1$ projective measurement $\{|j\rangle \langle j|\}_{j=1}^{d}$. Recently, Bischof, Kampermann and Bru\ss\  \cite{PRL2019,PRA2021} generalized the conventional framework of coherence to the case of general positive operator-valued measures (POVMs), by replacing the projective measurements with POVMs. Recently, several POVM-based coherence measures have been proposed \cite{PRL2019,PRA2021,Xu2020}, such as relative entropy of POVM-based coherence $C_{r}(\rho ,E)$, $l_{1}$-norm of POVM-based coherence $C_{l_{1}}(\rho ,E)$, robustness of POVM-based coherence $C_{rob}(\rho,E)$, and POVM-based coherent based on Tsallis relative entropy $C_{T, \lambda}(\rho,E)$. Moreover, maximum relative entropy of coherence for quantum channels has been introduced in \cite{Jin2021}.

Similar to the case of quantum entanglement, the superposition of two incoherent states may give rise to a maximally coherent state, for instance, $|\Omega_1\rangle=\frac{1}{\sqrt{2}}(|0\rangle+|1\rangle)$ in the computational basis $|0\rangle,|1\rangle$. While the superposition of two maximally coherent states may produce an incoherent state, for example, $| \Omega_2 \rangle = \frac{1}{\sqrt{2}} (|+ \rangle +|- \rangle)$, where $|\pm\rangle= \frac{1}{\sqrt{2}} (|0 \rangle \pm |1 \rangle)$.
In \cite{liyue,Li2017,Yue2016,shao2019} the coherence for superposition states of two mutual orthogonal states has been discussed.

In this work, we investigate the POVM-based coherence of superposition states. More specifically, given a superposition state,
\begin{eqnarray}
|\Omega \rangle=\sum_{k=1}^{n} \alpha_k | \phi_k \rangle,
\end{eqnarray}
we explore the relationship between the POVM-based coherence of the superposition state $|\Omega \rangle$ and the POVM-based coherence of the states $|\phi_k\rangle$ in the superposition. We focus on four POVM-based coherence measures $C_{r}(\rho,E)$, $C_{l_{1}}(\rho,E)$, $C_{rob}(\rho,E)$, and $C_{T, \lambda}(\rho,E)$ and present the upper and lower bounds of the inequalities satisfied by these coherence measures. We illustrate our results by detailed single and two-qubit examples.

\section{POVM-based coherence of superposition states}
In order to quantitatively describe the coherence resources contained in a given quantum state, a variety of coherence measures has been introduced from different perspectives. Before giving our main results, we first recall the definitions of $C_{r}(\rho,E)$, $C_{l_{1}}(\rho,E)$, $C_{rob}(\rho,E)$ and $C_{T, \lambda}(\rho,E)$ with respect to a positive operator-valued measure given by $d$ measurement operators $E=\{E_j\geq0\}_{j=1}^{d}$, $\sum_{j}E_{j}=I$ with $I$ the identity operator.

\subsection{The POVM-based coherence measures}
The relative entropy coherence measure is a commonly used and well-defined coherence measure \cite{2014prl}, which is tightly related to the optimal distillation rate in standard coherent distillation \cite{Winter2016} and the minimum amount of noise for complete decoherence \cite{Plenio2017,Singh2015}. Given a POVM $E$, the relative entropy POVM-based coherence is defined by \cite{PRL2019,PRA2021}
\begin{eqnarray}\nonumber
C_{r}(\rho, E)=\sum^d_{j=1}S(\sqrt{E_{j}}\rho \sqrt{E_{j}})-S(\rho ),
\end{eqnarray}
where $S(X)=-\textrm{Tr}(X\log X)$ is the von Neumann entropy for positive semidefinite matrix $X$. For a pure state $|\phi \rangle$, $C_{r}(\rho,E)$ can be expressed as
\begin{eqnarray}
C_{r}(\phi, E)=\sum^d_{j=1}S(\sqrt{E_{j}}| \phi  \rangle  \langle  \phi | \sqrt{E_{j}}). \label{eq211}
\end{eqnarray}

The $l_{1}$-norm POVM-based coherence $C_{l_{1}}(\rho ,E)$ of a density matrix $\rho$ is defined by \cite{PRA2021,Xu2020},
\begin{eqnarray}\nonumber
C_{l_{1}}(\rho ,E)=\sum^d_{i\neq j =1}||\sqrt{E_{i}}\rho \sqrt{E_{j}}||_{\textrm{tr}},
\end{eqnarray}
with $||X||_{\textrm{tr}}=\textrm{Tr}\sqrt{X^{\dag }X}$ the trace norm of matrix $X$. For a pure state $| \phi \rangle$, the $l_{1}$-norm POVM-based coherence is given by
\begin{eqnarray}
C_{l_{1}}(\phi ,E)=\sum^d_{i\neq j=1} \textrm{Tr} |\sqrt{E_{i}}|\phi  \rangle  \langle  \phi | \sqrt{E_{j}}|.\label{eq311}
\end{eqnarray}

The robust coherence is closely related to coherence sightings and can be used to quantify the advantages of quantum states in phase discrimination tasks. The
robust POVM-based coherence $C_{rob}(\rho,E)$ can be expressed as \cite{PRA2021},
\begin{eqnarray}\nonumber
C_{rob}(\rho,E)=\min_{\tau \in S^{'}} \left\lbrace s \geq0 : s \tau_{i,j}=-\sqrt{E_{i}}\rho \sqrt{E_{j}}, \forall i \neq j \right\rbrace.
\end{eqnarray}
where $\tau=\sum_{i,j}\tau_{i,j} \otimes |i \rangle  \langle j|$ as $\tau=\sum_{i,j}|\phi_i \rangle  \langle  \phi_j |  \otimes |i \rangle  \langle j| +\sum_{i}(I-|\phi_i \rangle  \langle  \phi_i | ) \otimes |i \rangle  \langle i|$.
Note that for pure states $|\phi\rangle$, the robust POVM-based coherence $C_{rob}(\rho,E)$ and the $l_{1}$-norm POVM-based coherence satisfy the following relation,
\begin{eqnarray}\nonumber
C_{rob}(\phi,E)=C_{l_{1}}(\phi ,E).
\end{eqnarray}

The Tsallis relative entropy of POVM-based coherence $C_{T,\lambda }(\rho ,E)$ introduced in \cite{Xu2020} is given by
\begin{equation}
C_{T,\lambda}(\rho,E)=\frac{1}{\lambda-1}\Big\{\sum^d_{j=1}\textrm{Tr}[(\sqrt{E_{j}}\rho^{\lambda} \sqrt{E_{j}})^{1/\lambda }]-1\Big\}, \label{eq111}
\end{equation}
for $\lambda \in (0,1)\cup (1,2]$.

For further use, we extend the above definition of the POVM-based coherence of a pure state to the coherence of two pure states, by analog to the case of quantum entanglement \cite{Con2011}.
Let $|\phi\rangle$ and $|\psi\rangle$ be two pure states.
We define $C_{l_{1}}(\phi,\psi)$ and $C_{T,\lambda}(\phi,\psi)$ as the POVM coherence of $|\phi\rangle$ and $|\psi\rangle$ by
\begin{align}
C_{l_{1}}(\phi,\psi ,E)=\sum^d_{i\neq j=1}\textrm{Tr}|\sqrt{E_{i}}| \phi \rangle \langle \psi | \sqrt{E_{j}}|
\end{align}
and
\begin{align}
C_{T,\lambda}(\phi,\psi ,E)
=\frac{1}{\lambda-1} \Big\{\sum_{j=1}^{d}\textrm{Tr}[(\sqrt{E_{j}}(| \phi \rangle \langle \psi | )^{\lambda} \sqrt{E_{j}})^{1/\lambda }]-1 \Big\}. \nonumber
\end{align}
Notice that according to the above definitions, we have $C(\phi,\phi, E)=C(\phi, E)$.

\subsection{The upper and lower bounds of POVM-based coherence measures}
In the following, we use the relative entropy POVM-based coherence to investigate the relationship between the coherence of two arbitrary pure states and the POVM-based coherence of their superposed state.

Before we present our two main results (Theorem 1), we briefly review the properties of the von-Neumann entropy $S(\rho)$ in~\cite{NiBook,Linden2006} and provide a slight improvement of it. The authors in~\cite{Linden2006} have used two properties of $S(\rho)$:
\begin{equation}
|\alpha_1|^2S(\rho)+|\alpha_2|^2S(\sigma)\leq  S(|\alpha_1|^2\rho+|\alpha_2|^2\sigma)
\label{concave}
\end{equation}
and
\begin{eqnarray*}
S(|\alpha_1|^2\rho+|\alpha_2|^2\sigma)\leq |\alpha_1|^2S(\rho)+|\alpha_2|^2S(\sigma)+h_2(|\alpha_1|^2).
\end{eqnarray*}
where $h_2(x) = - x \log x - (1 - x)\log (1 -x)$ and $\left| \alpha_1 \right|^2  + \left|\alpha_2\right|^2 = 1$.
Now, consider $| \Omega \rangle =\alpha_1 | \phi  \rangle  +\alpha_2 | \psi \rangle$, we obtain
\begin{eqnarray*}
S ( \sqrt{E_{j}} | \Omega \rangle \langle \Omega |  \sqrt{E_{j}} )
\leq| \alpha_1 |^2 S ( \sqrt{E_{j}} | \phi \rangle \langle \phi | \sqrt{E_{j}} )+ |\alpha_2|^2 S ( \sqrt{E_{j}} | \psi  \rangle \langle \psi | \sqrt{E_{j}} ) + h_2( | \alpha_1 |^2 ).
\end{eqnarray*}
Using Eq.~(\ref{eq211}), we get,
\begin{eqnarray}
C_{r}( \Omega , E) \leq |\alpha_1 |^2 C_{r} ( \phi, E)+|\alpha_2|^2 C_{r} (\psi , E) + h_2( {|\alpha_1|^2 } ), \label{relative1}
\end{eqnarray}

Based on the above results, we have the following conclusion.

\begin{theorem}
Given two different pure states $| \phi  \rangle $ and $ | \psi  \rangle$, denote
$|\Omega\rangle=\alpha|\phi\rangle+\beta|\psi\rangle$, where $\alpha$ and $\beta$ are complex numbers. The relative entropy POVM-based coherence $C_{r}(\Omega^{'},E)$ of the state $|\Omega^{'}\rangle=\frac{|\Omega\rangle}{||\Omega||}$, with
$||\Omega||=|\langle\Omega|\Omega\rangle|$ the normalization constant, has an upper bound
\begin{eqnarray}
p_1[\mu C_{r}(\phi,E)+(1-\mu)C_{r}(\psi,E)+h_2(\mu)],\label{eq231}
\end{eqnarray}
and a lower bound $L=\max\{L_1,L_2,0\}$ with
\begin{align}\label{eq711}
&L_1=p_2 C_{r}( \phi  ,E)-\frac{1-\nu}{\nu} C_{r}( \psi, E)-\frac{1}{\nu} h_2 (\nu),\nonumber \\
&L_2=p_3 C_{r}( \psi  ,E)-\frac{1-\xi}{\xi} C_{r}( \phi, E)-\frac{1}{\xi} h_2 (\xi),\nonumber
\end{align}
where $h_2( x ) = - x\log x - (1 - x)\log (1 - x)$,
\begin{align}
&p_1=\frac{(1-\mu) |\alpha|^2 +\mu |\beta|^2}{\mu(1-\mu)||\Omega||^2 },
~~\mu=\frac{|\alpha|^2}{\cos^2 \theta},\nonumber\\
&p_2=\frac{(1-\nu) |\alpha|^2}{(1-\nu)||\Omega||^2 +\nu |\beta|^2},
~~\nu=\frac{\sin^2\theta ||\Omega||^2}{\sin^2\theta ||\Omega||^2 +|\beta|^2\cos^2\theta},\nonumber\\
&p_3=\frac{(1-\xi) |\beta|^2}{(1-\xi)||\Omega||^2+\xi |\alpha|^2},
~~\xi=\frac{\sin^2\theta ||\Omega||^2}{\sin^2\theta ||\Omega||^2 +|\alpha|^2\cos^2\theta}\nonumber
\end{align}
for $0<\mu,\nu,\xi<1$ and
$\frac{|\alpha|^2}{\cos^2 \theta}+\frac{|\beta|^2}{\sin^2 \theta}=1$.
\end{theorem}

{\sf Proof} We prove the theorem by introducing an auxiliary system.
First we prove the upper bound (\ref{eq231}).
Consider the following bipartite state in systems $A$ and $B$,
\begin{eqnarray}\nonumber
|\chi \rangle^{AB} =\sqrt{\mu} | \phi \rangle^A |0 \rangle ^B
+\sqrt{1-\mu} | \psi \rangle ^A |1 \rangle^B.
\end{eqnarray}
According to (\ref{relative1}), the relative entropy POVM-based coherence of $| \chi \rangle$ can be expressed as
\begin{eqnarray}\nonumber
C_{r}( \chi,E) \leq \mu C_{r}( \phi  ,E)+ (1-\mu) C_{r}( \psi, E)+h_2 (\mu).
\end{eqnarray}
Measuring the ancillary system $B$ with Kraus operators
\begin{eqnarray}
&K_1=|0\rangle(\cos \theta e^{i \omega_1} \langle 0|+\sin \theta e^{i \omega_2} \langle 1 |), \nonumber\\
&K_2=|1 \rangle(-\sin \theta e^{-i \omega_2} \langle 0 | + \cos \theta e^{- i \omega_1} \langle 1|) , \nonumber
\end{eqnarray}
one gets the collapsed state
\begin{eqnarray}\nonumber
|\chi_1\rangle|0\rangle=\left(\sqrt{\frac{\mu}{p}}\cos\theta e^{i\omega_1}|\phi\rangle+\sqrt{\frac{1-\mu}{p}}\sin\theta e^{i\omega_2}|\psi\rangle\right)|0\rangle,
\end{eqnarray}
with probability
\begin{align}
p=\|\sqrt{\mu} \cos \theta e^{i \omega_1} | \phi \rangle + \sqrt{1-\mu } \sin \theta e^{i \omega_2}| \psi \rangle\|^2,
\end{align}
and the collapsed state
\begin{eqnarray}\nonumber
| \chi_2 \rangle | 1 \rangle =\left( \sqrt{\frac{1-\mu}{1-p} }\cos\theta e^{-i \omega_1}| \psi \rangle- \sqrt{\frac{\mu}{1-p}}\sin \theta e^{i \omega_2} | \phi \rangle \right) |1 \rangle,
\end{eqnarray}
with probability $1-p$. As the measurement can be seen as an incoherent operation, we obtain the following inequality,
\begin{eqnarray}\nonumber
p C_{r}( \chi_1, E)+ (1-p) C_{r}( \chi_2, E)\leq \mu C_{r}( \phi, E)+ (1-\mu) C_{r}( \psi, E)+h_2 (\mu).
\end{eqnarray}
Since $C_{r}( \chi_2, E) \geq 0$, we have
\begin{eqnarray}\label{eq132}
C_{r}( \chi_1, E) \leq \frac{1}{p} [\mu C_{r}( \phi, E)+ (1-\mu) C_{r}( \psi, E)+h_2 (\mu)].
\end{eqnarray}

Now we set $| \chi_1 \rangle=| \Omega^{'} \rangle$. We get
\begin{eqnarray}
\frac{\alpha}{||\Omega||}=\sqrt{\frac{\mu}{p}}\cos \theta e^{i \omega_1},~~
\frac{\beta}{||\Omega||}=\sqrt{\frac{1-\mu}{p}}\sin \theta e^{i \omega_2}. \nonumber
\end{eqnarray}
It is straightforward to derive that
\begin{eqnarray}
p=\frac{\|\Omega\|^2\mu(1-\mu)}{(1-\mu)| \alpha|^2 +\mu| \beta|^2}. \label{eq133}
\end{eqnarray}
Substituting (\ref{eq133}) into (\ref{eq132}), we have
\begin{eqnarray}
C_{r}(\Omega^{'},E)\leq p_1 \left[\mu C_{r}( \phi,E)+ (1-\mu) C_{r}( \psi, E)+h_2 (\mu)\right],\nonumber
\end{eqnarray}
with $0<\mu<1$, $\mu=\frac{|\alpha|^2}{\cos^2 \theta}$ and $p_1=\frac{(1-\mu) |\alpha|^2 +\mu |\beta|^2}{\mu(1-\mu)\|\Omega\|^2}$.

To prove the lower bound, we can consider the following bipartite states,
\begin{eqnarray}
| \chi^{'} \rangle ^{AB} &=&\sqrt{\nu} |\Omega^{'} \rangle ^A |0 \rangle ^B +\sqrt{1-\nu} | \psi \rangle ^A |1 \rangle ^B, \nonumber\\
| \chi^{''} \rangle ^{AB} &=&\sqrt{\xi} |\Omega^{'} \rangle ^A |0 \rangle ^B +\sqrt{1-\xi} | \phi \rangle ^A |1 \rangle ^B, \nonumber
\end{eqnarray}
Similar to the proof of the upper bound (\ref{eq231}), one obtains easily the lower bound $L$.
$\Box$

Concerning the $l_{1}$-norm of the POVM-based coherence measure, we have the following conclusion.

\begin{theorem}
Given $| \Omega  \rangle = \sum^{n}_{k=1} \alpha_k |\phi_k \rangle$ with $\alpha_k$ complex numbers, the POVM-based coherence of the superposed sate $| \Omega^{'} \rangle=\frac{| \Omega \rangle}{||\Omega||}$ has an upper bound
\begin{eqnarray}
\|\Omega\|^{-2}\Bigg(\sum^{n}_{k=1}|\alpha_k|^2C_{l_{1}}(\phi_k,E)+M\sum^{n}_{k\neq k^{'}=1}| \alpha_k \alpha_{k^{'}}|\Bigg),\label{eq313}
\end{eqnarray}
and a lower bound
\begin{eqnarray}
\|\Omega\|^{-2}\Bigg(\sum^n_{k=1}| \alpha_k|^2 C_{l_{1}}( \phi_k, E)-M\sum^n_{k\neq k^{'}=1}| \alpha_k \alpha_{k^{'}}|\Bigg),\label{q3}
\end{eqnarray}
where $M=(d-1)\sum^d_{i=1}\Big\|\sqrt{E_{i}} |\phi_k \rangle \langle \phi_{k^{'}}|\Big\|_{\textrm{tr}}$.
\end{theorem}

{\sf Proof }
From the definition of the $l_{1}$-norm POVM-based coherence (\ref{eq311}), we have
\begin{align}
\|\Omega\|^2 C_{l_{1}}(\Omega^{'},E)&=\sum^d_{i\neq j=1} \text{Tr} \Big|\sqrt{E_{i}}\sum^n_{kk^{'}=1}\alpha_k\alpha_{k^{'}}
|\phi_k\rangle\langle\phi_{k^{'}}|\sqrt{E_{j}}\Big|\nonumber\\
&\leq \sum^n_{k=1} |\alpha_k |^2\sum^d_{i\neq j=1}\text{Tr}\Big|\sqrt{E_{i}} |\phi_k \rangle \langle \phi_{k}|\sqrt{E_{j}}\Bigg| +\sum^n_{k\neq k^{'}=1}|\alpha_k\alpha_{k^{'}}|\sum^d_{i\neq j=1}\text{Tr}\Big|\sqrt{E_{i}}|\phi_k\rangle\langle\phi_{k^{'}}|\sqrt{E_{j}}\Big|  \nonumber\\
&= \sum^n_{k=1} |\alpha_k |^2 C_{l_{1}}( \phi_k, E) +\sum^n_{k \neq k^{'}=1} |\alpha_k \alpha_{k^{'}}| C_{l_{1}}( \phi_k, \phi_{k^{'}}, E) \nonumber\\
&\leq \sum^n_{k=1} |\alpha_k |^2 C_{l_{1}}( \phi_k, E)+ \sum^n_{k\neq k^{'}=1}| \alpha_k \alpha_{k^{'}}| M, \nonumber
\end{align}
where the first inequality is due to $|a+b| \leq |a| +|b|$, the second inequality is based on the Theorem 2 in Ref. \cite{xu2022} with $M=(d-1)\sum^d_{i=1}\Big\|\sqrt{E_{i}} |\phi_k \rangle \langle \phi_{k^{'}}|\Big\|_{\textrm{tr}}$.

Next, by using the inverse triangle inequality, $|a+b| \geq ||a|-|b|| \geq |a|-|b|$, similar to the proof of the upper bound we can easily get (\ref{q3}).
$\Box$

The robustness of coherence \cite{piani2016, PRA2021} plays an important role in the characterization of quantum states in phase discrimination. For pure states, the robustness of POVM-based coherence is equivalent to the $l_{1}$-norm of POVM-based coherence. Therefore, Theorem 2 also gives rise to the bounds for the robustness of POVM-based coherence.

Concerning the POVM-based coherence of Tsallis relative entropy $C_{T,\lambda }(\rho ,E)$, we present the following lemma.
\begin{lemma}
For $\lambda \in (0,1)\cup (1,2]$, we have
\begin{eqnarray}
C_{T,\lambda }(\rho, E)\leq -\ln_{\lambda}\Bigg(\frac{1}{[ d\sum^d_{j=1}\textrm{Tr}(\sqrt{E_{j}}\rho^{2} \sqrt{E_{j}})]^{1/\lambda}}\Bigg),\nonumber
\end{eqnarray}
where $\ln_{\lambda}x=\frac{x^{1-\lambda}-1}{1-\lambda}$.
\end{lemma}

{\bf Proof.} Due to (\ref{eq111}), for $\lambda \in (0,1)\cup (1,2]$ according to  \cite{pra2016, mubs}, we obtain
\begin{eqnarray}
C_{T,\lambda }(\rho, E)\leq\frac{[d^{\lambda-1} \sum^d_{j=1}\textrm{Tr}(\sqrt{E_{j}}\rho^{\lambda}
\sqrt{E_{j}})]^{1/\lambda}-1}{\lambda-1}.
\label{ubcma}
\end{eqnarray}
Applying the Jensen's inequality, we have
\begin{eqnarray}
\frac{[\sum^d_{j=1} \textrm{Tr}(\sqrt{E_{j}}\rho^{\lambda} \sqrt{E_{j}})]^{1/\lambda}}{\lambda-1}\leq
\frac{[( \sum^d_{j=1} \textrm{Tr}(\sqrt{E_{j}}\rho^{2} \sqrt{E_{j}}))^{1/\lambda}]^{\lambda-1}}{\lambda-1}.
\label{jencor}
\end{eqnarray}
Combining (\ref{jencor}) and (\ref{ubcma}), we complete the proof. $\Box$

Based on the above results, we have the following conclusion.

\begin{theorem}
Given a state $|\Omega\rangle =\sum^n_{k=1}\alpha_k|\phi_k\rangle$ with $\alpha_k$ the complex numbers. The POVM-based coherence of the normalized superposed state
$|\Omega^{'} \rangle=\frac{|\Omega\rangle}{\|\Omega\|}$ has an upper bound
\begin{align}\label{eq112}
\|\Omega\|^{-2}\Big(\sum^n_{k=1}|\alpha_k|^2C_{T,\lambda}(\phi_k,E)-\sum^n_{k\neq k^{'}=1}|\alpha_k\alpha_{k^{'}}|\ln_{\lambda}X\Big)+\frac{1}{\|\Omega\|^{2}(\lambda -1)}\Big(\sum^n_{k=1}|\alpha_k|^2+\sum^n_{k\neq k^{'}=1} |\alpha_k\alpha_{k^{'}}|-\|\Omega\|^2\Big)\nonumber
\end{align}
for $\lambda\in (0,1)\cup(1,2]$, where
\begin{align}
X=1/ \Big( d\sum^d_{j=1}\textrm{Tr}[\sqrt{E_{j}}(|\phi_k \rangle \langle \phi_{k^{'}}| )^{2} \sqrt{E_{j}}]\Big)^{\frac{1}{\lambda}},
\end{align}
and a lower bound $L=\max\{L_{1},0\}$ for $\lambda \in (1,2]$, where
\begin{align}
\|\Omega\|^{2}L_{1}=\sum^n_{k=1}|\alpha_k|^2C_{T,\lambda}(\phi_k,E)+N\sum^n_{k\neq k^{'}=1}|\alpha_k \alpha_{k^{'}}|+\frac{1}{(\lambda-1)}\Big(\sum^n_{k=1}|\alpha_k|^2+\sum^n_{k\neq k^{'}=1}|\alpha_k\alpha_{k^{'}}|-\|\Omega\|^2\Big)\nonumber
\end{align}
and
\begin{align}
N=\frac{1}{\lambda-1}\Big\{\sum^d_{j=1}\textrm{Tr}[(\sqrt{E_{j}})^{\frac{1}{\lambda}} |\phi_k \rangle \langle \phi_{k^{'}}| (\sqrt{E_{j}})^{\frac{1}{\lambda}}]-1\Big\}.\nonumber
\end{align}
\end{theorem}

{\sf Proof}
Let $\{|i\rangle_{i=1}^m\}$ be a set basic vectors such that $|\phi_k\rangle=\sum_{i = 1}^m b^k_i|i\rangle$. We have $|\Omega\rangle=\sum^n_{k=1}\sum^m_{i}\alpha_k b^k_i|i\rangle$ and
\begin{widetext}\begin{align}
\|\Omega\|^2 C_{T,\lambda}(\Omega^{'},E)
=&\frac{1}{\lambda -1}\Bigg\{\sum^d_{j=1}\textrm{Tr}\Bigg[\sqrt{E_{j}}\Big[\sum^n_{k k^{'}=1}\sum^m_{i i^{'}=1} \alpha_k b^k_i \alpha_{k^{'}} b^{k^{'}}_{i^{'}}|i\rangle \langle i^{'}|\Big]^{\lambda }\sqrt{E_{j}}\Bigg]^{\frac{1}{\lambda}}-\|\Omega\|^2\Bigg\}\nonumber\\
=&\frac{1}{\lambda -1}\Bigg\{\sum^n_{k=1} |\alpha_k|^2 \sum^d_{j=1}\textrm{Tr}\Bigg[\sqrt{E_{j}}\Big[\sum^m_{i i^{'}=1} |b^k_i b^k_{i^{'}}|\cdot|i\rangle\langle i^{'} |\Big]^{\lambda} \sqrt{E_{j}}\Bigg]^{\frac{1}{\lambda}}-\|\Omega\|^2\nonumber\\
&+\sum^n_{k\neq k^{'}=1} |\alpha_k \alpha_k^{'}|
\sum^d_{j=1}\textrm{Tr}\Bigg[\sqrt{E_{j}}\Big[\sum^m_{i i^{'}=1}|b^k_i b^{ k^{'}}_{i^{'}}|\cdot|i\rangle\langle i^{'} |\Big]^{\lambda}\sqrt{E_{j}}\Bigg]^{\frac{1}{\lambda}}\Bigg\}\nonumber\\
=&\sum^n_{k=1} |\alpha_k|^2 C_{T,\lambda }(\phi_k, E)+\sum^n_{k\neq k^{'}=1}|\alpha_k \alpha_{k^{'}}| C_{T,\lambda }(\phi_k,\phi_{k^{'}},E)\nonumber\\
&+\frac{1}{\lambda -1}\Big(\sum^n_{k=1}| \alpha_k|^2+ \sum^n_{k\neq k^{'}=1}|\alpha_k \alpha_{k^{'}}|-\|\Omega\|^2\Big).\nonumber
\end{align}\end{widetext}
Based on the Lemma 1, we obtain the upper bound in theorem.

Next, by using the Araki-Lieb-Thirring Inequality \cite{ALT2008}, $\textrm{Tr}(A^{r}B^{r}A^{r})^{q}\geq\textrm{Tr}(ABA)^{rq}$ for $r\geq 1$ and $q\geq 0$,
we have
\begin{eqnarray}
C_{T,\lambda }( \phi_k,\phi_{k^{'}},E)
&&=\frac{1}{\lambda -1}\Bigg\{\sum^d_{j=1}\textrm{Tr}\Big[(E_{j}^{\frac{1}{2\lambda}})^{\lambda} (|\phi_k \rangle \langle \phi_k^{'}|)^{\lambda }(E_{j}^{\frac{1}{2\lambda}})^{\lambda}\Big]^{1/\lambda }-1\Bigg\}\nonumber\\
&&\geq \frac{1}{\lambda -1}\Bigg\{\sum^d_{j=1}\textrm{Tr}\Big[(\sqrt{E_{j}})^{1/\lambda }|\phi_k \rangle \langle \phi_k^{'}| (\sqrt{E_{j}})^{1/\lambda }\Big]-1\Bigg\},\nonumber
\end{eqnarray}
where $\lambda \in (1,2]$.
Then, we obtain the lower bound in theorem.
$\Box$

\section{Numerical results}
In this section, we demonstrate numerically our results by investigating different superposition states. We consider a single-qubit case for the $l_1$ coherence measure and two-qubit case for the relative entropy and the Tsallis relative entropy coherence measures. In both cases, we plot the relations between exact value, upper bound, and lower bound. Crucially, a POVM of $n$-qubit system is expressed as $4^{n}$ linear independent positive operators $\{E_{i}=A_{i}A_{i}^{\dag}\}_{i=0}^{4^{n}-1}$, which can be obtained according to the Naimark theorem \cite{Peres2006Quantum,Decker2005Implementation}. Specially, each measurement operator $A_{i}$ embedded in a larger unitary can be found via a projective measurement in the standard computational basis of a larger Hilbert space.

Consider a POVM with respect to a single-qubit system $\{E_{i}=A_{i}A_{i}^{\dag}\}_{i=0}^{3}$, which can be realized by performing a two-qubit unitary $U=\sum_{ijkl=0}^{3}u_{ij}^{kl}|ij\rangle\langle kl|$ on the qubit system and an ancillary system. Implementing $U$ on an initial state $\rho_{a}\otimes|0\rangle\langle0|_{b}$, one measures the two qubits system in the standard computational basis $\{|m\rangle\}_{m=0}^{3}$. Denote $(q_a,q_b)$ the measurement outcome with $q_a,q_b\in\{0,1\}$. The corresponding probability of each outcome is given by
\begin{align}
P_{(q_a,q_b)}&=\langle q_aq_b|U(\rho_{a}\otimes|0\rangle\langle0|_{b})U^{\dag}|q_aq_b\rangle\nonumber\\
&=\sum_{kl=0}^{3}u_{q_aq_b}^{kl}\langle kl|(\rho_{a}\otimes|0\rangle\langle0|_{b})\sum_{kl=0}^{3}(u_{q_aq_b}^{kl})^{*}|kl\rangle\nonumber\\
&=\sum_{k=0}^{3}u_{q_aq_b}^{k0}\langle k|\rho_{a}\sum_{k=0}^{3}(u_{q_aq_b}^{k0})^{*}|k\rangle\nonumber\\
&=\textrm{Tr}[|E_{q_aq_b}\rangle\langle E_{q_aq_b}|\rho_{a}],
\end{align}
where $|E_{(q_aq_b)}\rangle=\sum_{k=0}^{3}u_{q_aq_b}^{k0}|k\rangle$. Hence, the measurement operators of each POVM is given by $E_{i}=E_{(q_aq_b)}=|E_{(q_aq_b)}\rangle\langle E_{(q_aq_b)}|$ for $i=0,\cdots,3$, where $i$ has a binary representation $i=2q_a+q_b$. In particular, let us we consider the following two-qubit unitary $U$,
\begin{align}
U(\boldsymbol{\theta})=\textrm{CNOT}\cdot\Big[R_{y}(\theta_1)\otimes R_{y}(\theta_2)\Big],
\end{align}
parameterized by $\boldsymbol{\theta}=(\theta_{1},\theta_{2})$, where $R_{y}(\theta_j)=e^{-\imath\theta_j\sigma_{y}/2}$, $\imath^{2}=-1$, and \textrm{CNOT} is the CNOT-gate. Setting $\boldsymbol{\theta}=(0.301723,0.011681)$ we have a POVM given by
\begin{align}
&|E_{0}\rangle=(u_{00}^{00},u_{00}^{10})^{\dag},~~
|E_{1}\rangle=(u_{01}^{00},u_{01}^{10})^{\dag},\nonumber\\
&|E_{2}\rangle=(u_{10}^{00},u_{10}^{10})^{\dag},~~
|E_{3}\rangle=(u_{11}^{00},u_{11}^{10})^{\dag}.\nonumber
\end{align}

\begin{figure}[]
\includegraphics[scale=1.2]{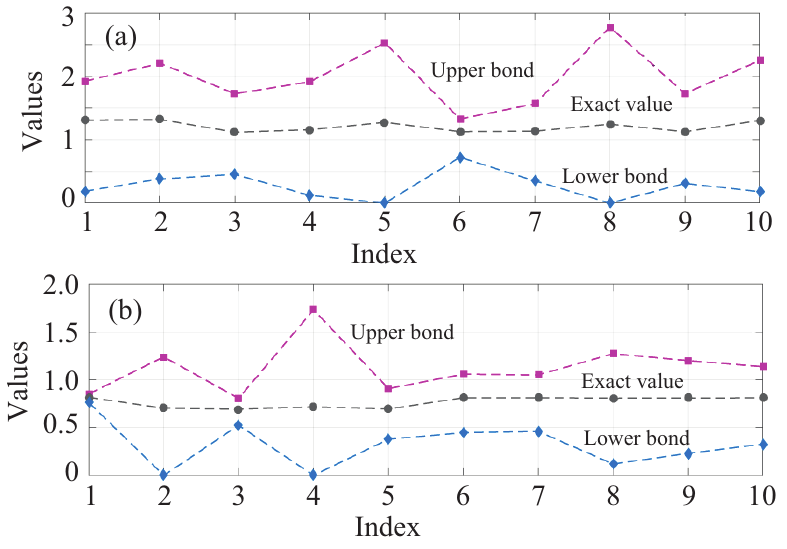}
\caption{(a) The relative entropy of POVM-based coherence. (b) The $l_{1}$ of POVM-based coherence.}
\label{Fig1}
\end{figure}

Given two $1$-qubit states $|\phi\rangle=e^{-\imath\theta_j0.432/2}|0\rangle$ and $|\psi\rangle=e^{-\imath\theta_j0.618/2}|0\rangle$, we calculate the exact value, upper bound, and lower bound of $l_1$ coherence of the superposition state
\begin{align}
|\Omega\rangle=\alpha|\phi\rangle+\beta|\psi\rangle,~~|\Omega^{'}\rangle
=\frac{|\Omega\rangle}{\|\Omega \|},
\end{align}
in which the coefficients $\alpha,\beta$ are random scalar drawn from the uniform distribution in the interval $(0,1)$. We run our procedure $10$ times by randomly choosing the coefficients $\alpha,\beta$. Fig. (\ref{Fig1}.b) shows the upper and lower bounds of $C_{l_{1}}(|\Omega^{'}\rangle\langle\Omega^{'}|,E)$.

For two-qubit case, consider POVMs with $16$ measurement operators, $\{E_{i}=A_{i}A_{i}^{\dag}\}_{i=0}^{15}$, which can be realized by performing a unitary
$V=\sum_{\boldsymbol{i}\boldsymbol{j}=0}^{3}v_{i_1i_2i_3i_4}^{j_1j_2j_3j_4}
|\boldsymbol{i}\rangle\langle\boldsymbol{j}|$
on a four-qubit system, where the index $\boldsymbol{i}$ has binary representation $\boldsymbol{i}=i_1i_2i_3i_4$. Similar to the single qubit case, it is easily verified that each POVM has a form $E_{i}=|E_{(q_aq_bq_cq_d)}\rangle\langle E_{(q_aq_bq_cq_d)}|$ where $(q_aq_bq_cq_d)$ denote the measurement outcomes, $i=2^3q_a+2^2q_b+2q_c+q_d$, and
\begin{align}
|E_{(q_aq_bq_cq_d)}\rangle=\sum_{j_1j_2=0}^{1}v_{q_aq_bq_cq_d}^{j_1j_200}|j_1j_2\rangle.
\end{align}
Let us consider
\begin{align}
V(\boldsymbol{\gamma})=\prod_{i,j=0}^{2}\textrm{CNOT}_{i,i+1}\cdot \Big[R_{y}(\gamma_1)\otimes R_{y}(\gamma_2)\otimes R_{y}(\gamma_3)\otimes R_{y}(\gamma_4)\Big],
\end{align}
given by the parameter vector $\boldsymbol{\gamma}=(\gamma_{1},\gamma_{2},\gamma_{3},\gamma_{4})
=(0.30173,0.01168,0.53991,0.09537,0.14651)$.

\begin{figure}[]
\includegraphics[scale=1.2]{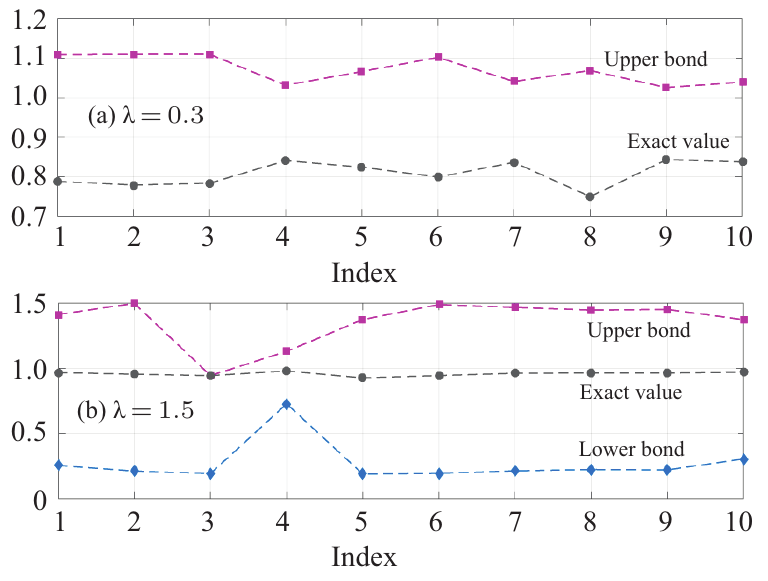}
\caption{The Tsallis relative entropy of POVM coherence.}
\label{Fig2}
\end{figure}

Given two $2$-qubit states
\begin{align}
|\phi_{1}\rangle=U(\boldsymbol{\theta}_{1})|00\rangle,~
|\psi_{1}\rangle=U(\boldsymbol{\theta}_{2})|00\rangle,~
|\psi_{2}\rangle=U(\boldsymbol{\theta}_{3})|00\rangle,\nonumber
\end{align}
where parameters $\boldsymbol{\theta}_{1}=(0.4827,0.3760)$, $\boldsymbol{\theta}_{2}=(0.9394,0.2212)$ and $\boldsymbol{\theta}_{3}=(0.1557,0.8190)$. We calculate the exact values, upper bounds, and lower bounds of relative entropy coherence of the superposition state
\begin{align}
|\Omega_{1}\rangle=\alpha|\phi_{1}\rangle+\beta|\psi_{1}\rangle,~~|\Omega_{1}^{'}\rangle
=\frac{|\Omega_{1}\rangle}{\|\Omega_{1}\|},
\end{align}
and Tsallis relative entropy coherence of the superposition state
\begin{align}
|\Omega_{2}\rangle=\alpha|\phi_{1}\rangle+\beta|\psi_{2}\rangle,~~|\Omega_{2}^{'}\rangle
=\frac{|\Omega_{2}\rangle}{\|\Omega_{2}\|},
\end{align}
in which the coefficients $\alpha,\beta$ are random scalar drawn from the uniform distribution in the interval $(0,1)$. We run our procedure $10$ times by randomly choosing the coefficients $\alpha,\beta$. Fig. (\ref{Fig1}a) shows the bounds of the relative entropy of POVM-based coherence. Only upper bound of the Tsallis relative entropy is shown in Fig. (\ref{Fig2}a) when $\lambda=0.3$. Fig. (\ref{Fig2}b) plots the upper and lower bounds of the Tsallis relative entropy of POVM coherence when the parameter $\lambda=1.5$. Clearly, the numerical results are good agree with our theoretical analysis.

\section{Conclusion}
The superposition principle of quantum states is fundamental in quantum mechanics.
Generally, quantum objects exhibit both the wave and particle nature. 
In particular, when the POVM $E$ is taken to be the von Neumann projective measurement, the coherence $C_{l_{1}}(\rho,E)$ stands for the wave property of quantum objects. In this case
the coherence $C_{l_{1}}(\rho,E)$ and the distinguishability
of which-path information (a measure of particle-property) in multi-path interference satisfy a trade-off relation \cite{roy}, which has been verified in a generalized multi-path delayed-choice experiment on a large-scale quantum nanophotonic chip \cite{wjw}.

Based on the POVM-based coherence measures $C_{r}(\rho,E)$, $C_{l_{1}}(\rho,E)$, $C_{rob}(\rho,E)$ and $C_{T,\lambda }(\rho, E)$, we haver studied coherence of superposition for arbitrary dimensional pure states. We have explored the relationship between the POVM-based coherence of a superposed state and the POVM-based coherence of the states in the superposition, and derived analytically the lower and upper bounds. To illustrate our results,
we have presented detailed examples related to the POVM-based $l_1$ norm coherence for the one-qubit case and the POVM-based relative entropy coherence for the two-qubit case.
Our results can be used to estimate the range of quantum coherence of superposed states, and may highlight investigations on superposition of coherence for quantum channels \cite{Jin2021}.

\bigskip
\noindent{\bf Acknowledgments}\, \, This work is supported by the National Natural
Science Foundation of China (NSFC) under Grants 12075159, 12171044 and 12175147; Beijing Natural Science Foundation (Grant No. Z190005); the Academician Innovation Platform of Hainan Province; Shenzhen Institute for Quantum Science and Engineering, Southern University of Science and Technology (No. SIQSE202001).

\end{document}